\documentclass[preprint,groupedaddress,floatfix,%
nofootinbib]{revtex4}

\usepackage{euscript}
\usepackage{dcolumn}
\usepackage{graphicx}
\usepackage{epsfig}
\usepackage{hyperref}
\usepackage{graphicx}
\usepackage{dcolumn}
\usepackage{longtable}
\usepackage{times}
\usepackage{amsmath,amssymb,bm}
\usepackage[english]{babel}
\usepackage[latin1]{inputenc}
\usepackage{times}
\usepackage[T1]{fontenc}
\usepackage{color}
\usepackage{fancybox}
\usepackage{sidecap}
\usepackage{graphicx}
\usepackage{epsfig}
\usepackage{psfrag}
\usepackage{bbm}
\usepackage[english]{babel}
\usepackage[latin1]{inputenc}
\usepackage{times}
\usepackage[T1]{fontenc}
\usepackage{color}
\usepackage{fancybox}
\usepackage{sidecap}
\usepackage{psfrag}
\usepackage{bbm}
\usepackage{wasysym}
\usepackage{slashed}
\usepackage{tikz}
\usepackage{euscript}

\newcommand{\mathi}{\mathrm{i}}
\newcommand{\mathe}{\mathrm{e}}

\newcommand{\nn}{\nonumber}

\begin{document}

\title{Simplicity condition and boundary-bulk duality}
\author{Fen Zuo\footnote{Email: \textsf{18802711412@139.com}}}
\affiliation{~}

\begin{abstract}
In the first-order formulation, general relativity could be formally viewed as the topological $BF$ theory with a specific constraint, the Plebanski constraint. $BF$ theory is expected to be the classical limit of the Crane-Yetter~(CY) topological state sum. In the Euclidean case, the Plebanski constraint could be lifted  in an elegant way to a quantum version in the CY state sum, called the simplicity condition. The constrained state sum is known as the Barrett-Crane~(BC) model. In this note we investigate this condition from the topological field theory side. Since the condition is in fact imposed on the faces, we want to understand it from the viewpoint of the surface theory. Essentially this condition could be thought of as resulting from the boundary-bulk dualtiy, or more precisely from the recent ``bulk=center'' proposal. In the language of topological phases, it corresponds to the diagonal anyon condensation, with the BC model being the condensed phase.

The CY state sum, and correspondingly the BC model, is usually constructed with a modular tensor category~(MTC) from representations of a quantum group at roots of unity. The category of representations of quantum group and that of modules of the corresponding affine Lie algebra are known to be equivalent as MTCs. This equivalence, together with the simplicity condition in the BC model, guarantees the construction of a full 2d rational conformal field theory through the Fuchs-Runkel-Schweigert formalism. We thus obtain a full-fledged 4d quantum geometry, which we name as ``stringy quantum geometry''. Some attracting features of such a new geometry are briefly discussed.

\end{abstract}
 \maketitle

\section{Introduction}
The essence of general relativity~(GR) is believed to be that it is established without any a-priori background. One would thus expect that the quantum theory must inherit this key property. Topological quantum field theory~(TQFT)~\cite{Segal:1987sk,Atiyah:1989vu} provides a natural framework for a background-independent formulation of quantum gravity~\cite{Baez:1995xq}. Indeed, GR has close relation to a specific topological theory, which is manifest in the first-order formalism. With the frame field $e$ and the spin connection $A$, the Einstein-Hilbert action could be compactly written as
\begin{equation}
S_{\text{GR}}(e,A)\sim \int_M (e\wedge e \wedge F),\label{eq.GR}
\end{equation}
where the wedge product is supposed to take over both the spacetime part and the internal part. Apparently, this could be viewed as the topological BF theory~\cite{Baez:1995ph,Baez:1999sr}:
\begin{equation}
S_{BF}(B,A)=\int_M (B\wedge F),\label{eq.BF}
\end{equation}
with the constraint
\begin{equation}
B=e\wedge e.\label{eq.Plebanski}
\end{equation}
This is essentially the Plebanski formulation for GR, so we may call (\ref{eq.Plebanski}) the Plebanski constraint. The simplification for the canonical quantization with the Ashtekar variable~\cite{Ashtekar:1986yd} indicates that we should also express the constraint through chiral variables. Then it becomes clear that the constraint simply says that the $B$ field must be a simple bivector, which is an equal-weight combination of the self-dual/anti self-dual parts~\cite{Barrett:1997gw}. Moreover, the simplicity of the bivector deduces that the two parts have the same norm~\cite{Barrett:1997gw}. This could then be lifted to a constraint in the quantum version of the BF-theory, namely the Crane-Yetter~(CY) construction~\cite{Crane:1993if}, at least in the Euclidean case. The resulting constrained state-sum defines the Barrett-Crane~(BC) model~\cite{Barrett:1997gw}.

Let us make this procedure more explicit for 4d Euclidean gravity as in~\cite{Barrett:1997gw}. Taking the spin cover Spin$(4)$ of the internal gauge group SO$(4)$, the self-dual and anti self-dual parts factorize as
\begin{equation}
\text{Spin}(4) \cong \text{SU}(2)\times \text{SU}(2).\label{eq.Spin}
\end{equation}
As in 3d~\cite{Ponzano-Regge}, in order to obtain finite state sums~\cite{Turaev:1992hq}, one has to deform the corresponding universal enveloping algebra to get the ``quantum'' group. The deformation parameter $q$ is supposed to be related to the cosmological constant~\cite{Baez:1995ph}. When $q$ takes special values at roots of unity, the category of irreducible representations of the quantum group possesses the nice structure of a ``modular tensor category''~(MTC)~\cite{Turaev:1992}, which could then be used to construct a CY state sum. In the BC model it turns out that one has to choose opposite braiding of the two quantum groups~\cite{Barrett:1997gw}. Denote $\EuScript{C}\equiv \text{Rep}(U_{q}sl_2)$ and $\EuScript{C}^-$ a copy of $\EuScript{C}$ but with reversed braiding. Then it means, here one has actually chosen the Deligne tensor product $\EuScript{C} \boxtimes \EuScript{C}^-$.
As a MTC, $\EuScript{C}$ contains only finitely many simple objects. Number the set of simple objects of $\EuScript{C}$ by $I$. The first part of the simplicity constraints, called the diagonal one, is imposed by selecting specific objects of the form $X_i\times X_i$, for any $i\in I$. Since in CY state sum one colors the triangles in the simplicial decomposition by the objects, the diagonal condition is essentially imposed on the surfaces. The non-diagonal part is related to the tetrahedra itself, or the dual four-valent vertices. A four-valent vertex could be split as two three-valent vertices, with one intermediate link. The three-valent vertices are colored with the intertwiners for $U_{q}sl_2$, which are completely fixed up to normalization. One is left with a single intermediate link. Then the non-diagonal condition is specified in the same way as the diagonal one. In this way one obtains a unified description for all the links, either the original ones dual to the triangles, or those coming from splitting. Such a unified description could be easily extended to general decomposition with arbitrary polyhedra~\cite{Yetter:98,Reisenberger:1998bn}. It is also vividly figured as ``left-handed area = right-handed area''~\cite{Reisenberger:1998fk}.

One could further simplify the above quantum condition a little more. The idea is similar to the Skein-theory description based on handle-decomposition~\cite{Roberts:1995}. Since in the CY state-sum only finite sums and products are involved, one could change the order of them arbitrarily. Also it is implicitly assumed that each allowed simple object appears in the final sum once and just once. Therefore one could color each link with a single special object:
\begin{equation}
A_{\text{cl}}\equiv \oplus_{i\in I} X_i\times X_i^\vee.\label{eq.simplicity}
\end{equation}
Here we have intensively replace the second factor by its dual, which does not make any difference since the representations are self-dual. Then the intertwiner for the individual simple objects must also be lifted to a specific intertwiner for three $A_{\text{cl}}$'s. This should in principle be possible, since all the simple objects are treated on the same footing. At the end the only thing that matters would be the normalization of the intertwiners, which we neglected here. In principle one should be able to give a rigorous proof of this following the procedure in~\cite{Roberts:1995}, which will not be pursued here. The main purpose of this paper is to explain the special nature of this object $A_{\text{cl}}$ from a holographic viewpoint. This will be the focus of the next section.

\section{``bulk=center''}

\subsection{``bulk=center''}
The CY construction is of the kind of ``absolute'' theory called in~\cite{Atiyah:1989vu}. In general we will need the ``relative'' theory~\cite{Atiyah:1989vu} or extended ones as advocated in~\cite{Baez:1995xq,Lurie:2009} using higher categories. Even these extended types still have two shortages: first, a TQFT constructed with some non-trivial higher category could turn out to be classical, which can not be justified directly from the category data; Secondly, although a formal description of holographic principle could be established~\cite{nLab-holography}, it is not clear how it is concretely realized in extended TQFTs. To overcome these difficulties, one needs to perform some kind of the state-operator correspondence, and to change the focus from the Hilbert spaces to the observables. From the physical point of view, the Hamiltonian formulation~\cite{Kitaev:1997wr,Levin:2004mi,Walker-Wang:2011} provides a nice solution to these problems. In particular, the Hamiltonian formulation allows for a concrete categorification of the holographic principle, which is encapsulated in the so-called ``bulk=center'' relation~\cite{Kong:2014qka,Kong:2015,Kong:2017}. While the original proposal is restricted to the gapped phases~\cite{Kong:2014qka,Kong:2015}~\footnote[1]{Another issue is the so-called perturbative gravitational anomaly in $4k+3$ spacetime dimensions~\cite{Kong:2014qka}. In $(2+1)d$ this would correspond to the framing anomaly~\cite{Witten:1988hf,Reshetikhin:1991}. This is expected to be resolved/restored~\cite{Witten:1988hf}\cite{Barrett:2004im,Walker-Wang:2011} when one lifts it to a $(3+1)d$ bulk theory, and thus will be neglected in this paper.}, a model-independent exposition is given recently~\cite{Kong:2017}~, which shows that the relation is universal~(which is also briefly mentioned in~\cite{Kong:2015}). In particular, a rigorous construction for the gapless edges of $(2+1)d$ bulk topological phases is made recently in~\cite{KZ:1705}, by generalizing the notion of enriched monoidal category~\cite{KZ:1704}. It is proved that the ``bulk=center'' relation is indeed valid in such a gapless case~\cite{KZ:1704}. In the following we shortly summarize their main idea here.

In the Hamiltonian formulation~\cite{Kitaev:1997wr,Levin:2004mi,Walker-Wang:2011} of TQFTs, the topological excitations of various codimensions could be identified as the physical observables. In $(n+1)$-dimension, this is compactly described by a unitary fusion $n$-category, which could also be viewed as a unitary $(n+1)$-category  with one simple object~\cite{Kong:2014qka,Kong:2015}. One could then study the relation between the theories in different spacetime dimensions. It is discovered that this could be elegantly captured by the mathematical notion of ``$Z_n$-center'' of a $(n-1)$ fusion category. Namely if a $(n+1)$-dimensional bulk theory is holographically/physically related to a $n$-dimensional boundary theory, the bulk excitations are given by the $Z_n$-center of the boundary excitations, which themselves are captured by a $(n-1)$ fusion category. In this rough statement we have somehow relaxed the required conditions specified in the original formulation~\cite{Kong:2014qka,Kong:2015}. But this does not cause any real problems since later we only apply it in the low-dimensional cases, which have been analyzed explicitly~\cite{Kong:2014qka}. We will not recall the rigorous definition of the $Z_n$-center, but only outline some properties. In the special case when $n=2$, it just coincides with the ordinary monoidal/Drinfel'd center $Z$. The $Z_n$-center was shown to possess an important property, namely the center of center is trivial. This can then be employed to define a generalized cohomology theory for the theories in different spacetime dimensions. A theory is called closed if the center of its excitations is trivial, otherwise it is anomalous; in other words, a closed boundary theory defines a trivial bulk theory, while an anomalous boundary necessarily gives rise to nontrivial bulk excitations and a bulk description is really necessary. A theory is called exact if its excitations are the center of the excitations of a boundary theory; in other words, it has a dual boundary description.

\subsection{4-3-2}~\footnote[2]{The title is borrowed from~\cite{Freed:2012}.}
Now we could revise the various topological state sums in physical space-time dimensions with the above holographic principle. This has essentially been done in the Hamiltonian approach, so we only need to translate them back to the ``path-integral'' formulation. Let us start with the previously mentioned CY state sum, which is expected to be the quantum version of the $BF$-theory. Originally it is constructed from a MTC~\footnote[3]{We are not going to talk about the more general construction with a premodular category~\cite{Crane:1994ji,Williamson:2016evv}, since they will not be directly related to the $BF$-theory anymore.}, e.g., Rep$(U_{q}sl_2)$ for $q$ at roots of unity, thus a special kind of $3$-category. One would then wonder if the CY construction satisfies the above boundary-bulk duality, so let's go to the observables. It turns out that the CY-theory is classical~\cite{Roberts:1995,Crane:1993cm}\cite{Walker-Wang:2011}, thus contains no non-trivial topological excitation at all. In other words, it is characterized by the trivial category $\bf{1}_4$. And if we want it to be dual to a boundary theory, the boundary theory must be closed, or anomaly free.

Since the state space of the $BF$-theory is given by the Chern-Simons~(CS) functional~\cite{Witten:1988hf}\cite{Smolin:1995vq}\cite{Baez:1995ph}, it is natural to expect that the two are holographically related. The mathematical formulation of the CS theory~\cite{Witten:1988hf} is believed to be the Reshetikhin-Turaev~(RT) TQFT~\cite{Reshetikhin:1991}, which could be further extended to give a $({3-2-1})$ extended theory. From the handle-decomposition description~\cite{Roberts:1995,Barrett:2004im}, it is clear that CY and RT theory are indeed directly related. In particular, the two could be constructed with exactly the same category data~(up to the framing anomaly mentioned previously), namely a MTC $\EuScript{C}_3$. However, no Hamiltonian formulation of RT theory has been realized, and thus it is hard to characterize it with the excitations. Since the RT theory assigns the objects of $\EuScript{C}_3$ to the circle $S^1$~\cite{Kirillov:2010nh,Turaev:2010,Balsam:2010,Balsam:2010B}, which is the lowest level in the extended theory, one may argue that the basic excitations are just specified by the category itself. Later we will see that this is indeed the case for a special class of MTCs. If we accept such an identification of excitations, the boundary-bulk duality asserts that the MTC must have a trivial center. In other words, its $Z_3$-center is trivial in the notion of~\cite{Kong:2014qka,Kong:2015},
\begin{equation}
Z_3(\EuScript{C}_3)=\bf{1}_4.
\end{equation}

In order to get a boundary description, we should further impose the exactness condition:
\begin{equation}
\EuScript{C}_3=Z_2(\EuScript{C}_2),
\end{equation}
for $\EuScript{C}_2$ a unitary fusion category~\footnote[4]{According to~\cite{Fuchs:2012dt}, this can be relaxed to a braided equivalence. Moreover, based on the notion of enriched monoidal categories, it has been proved that such an equivalence can be established for general MTCs~\cite{KZ:1704,KZ:1705}.}. As stated before, the $Z_2$-center coincides with the monoidal center of a fusion category. To simplify the notation, we will just write $\EuScript{C}$ instead of $\EuScript{C}_2$, and the monoidal center instead of the $Z_2$-center. Thus $\EuScript{C}_3=Z(\EuScript{C})$. Notice that the modular structure of $\EuScript{C}_3$ is automatically guaranteed when $\EuScript{C}$ is unitary fusion. Furthermore, if $\EuScript{C}$ is modular we have the following braided equivalence:
\begin{equation}
\EuScript{C}_3=Z(\EuScript{C})\cong \EuScript{C} \boxtimes \EuScript{C}^-,\label{eq.C2}
\end{equation}
where $\EuScript{C}^-$ is again obtained from $\EuScript{C}$ by reversing all the braiding. At this moment we only need its fusion structure. It is well known that from a unitary fusion category $\EuScript{C}$ one could construct the Turaev-Viro~(TV) state sum~\cite{Turaev:1992hq}, which is a convergent form of the Ponzano-Regge state sum for 3d quantum gravity~\cite{Ponzano-Regge}. The relation with RT invariants has been extensively studied~\cite{Roberts:1995,Barrett:2004im}. In~\cite{Kirillov:2010nh,Turaev:2010,Balsam:2010,Balsam:2010B} it is shown that it could be extended as a $(3-2-1)$ TQFT, and the extended theory is equivalent to RT-theory based on $Z(\EuScript{C})$:
\begin{equation}
Z_{\text{TV},\EuScript{C}}\cong Z_{\text{RT},Z(\EuScript{C})}.
\end{equation}

More importantly, we have an elegant Hamiltonian realization of TV-theory, the Levin-Wen model~\cite{Levin:2004mi}, which is constructed with the same data $\EuScript{C}$. A nice review of this model is given in~\cite{Walker-Wang:2011}. Roughly speaking, the model is built by labeling the edges in a 2d space lattice with objects of $\EuScript{C}$. The morphisms and associators are treated as operators acting on the lattice. Combining these operators one could construct a Hamiltonian, which essentially imposes ``gauge-invariant'' constraints on the vertices and ``flatness'' constraints on the plaquettes. The meaning of these constraints become more clear in an equivalent model, Kitaev's quantum double model~\cite{Kitaev:1997wr}, constructed with some weak Hopf algebra $H$ as the ``gauge group''.
The two are related according to the Tannaka-Krein reconstruction theorem~\cite{Joyal:1991}
\begin{equation}
\EuScript{C}\cong \text{Rep}(H).
\end{equation}
The equivalence of the extended versions of the two models, and with the extended TV/RT-theory, have been firmly established~\cite{Buerschaper:2009,Kadar:2009fs,Buerschaper:2010yf,Kirillov:2011}. Now with these Hamiltonian realizations, one could explicitly identify the various excitations, which appear whenever the constraints in the Hamiltonian are violated. It turns out they are characterized by the monoidal center $Z(\EuScript{C})$. Intuitively, this corresponds in the extended TV/RT theory to the data assigned to the circle $S^1$~\cite{Kirillov:2010nh,Balsam:2010,Balsam:2010B}. This also provides some evidence of our previous proposal that the excitations in general RT-theory could be identified with the constructing data $\EuScript{C}_3$.

The power of the Hamiltonian formulation is much beyond this. They could be further generalized to describe also the boundary theory, which are elaborated in~\cite{Kitaev-Kong:2012,Kong:2012ph}. Here a very nice notion, as a straightforward categorization of modules over rings, module categories over a monoidal category~\cite{Ostrik:2001}, perfectly fit in. As we mentioned, the Levin-Wen model is built with a 2d space lattice, with the edges colored with objects in $\EuScript{C}$. On the 1d boundary of the lattice, it is natural to color the boundary edge with objects of the module category ${}_\EuScript{C}\EuScript{M}$, and the vertices with the action morphisms. Moreover, the boundary excitations are specified by the monoidal category of endofunctors of ${}_\EuScript{C}\EuScript{M}$,
\begin{equation}
\EuScript{C}_\EuScript{M}^\vee:=\EuScript{F}\mathrm{un}_{\EuScript{C}}(\EuScript{M},\EuScript{M}).\nn
\end{equation}
It is shown that $\EuScript{C}$ and $\EuScript{C}_\EuScript{M}^\vee$ are Morita equivalent, which immediately leads to the boundary-bulk duality~\cite{Kitaev-Kong:2012}\cite{Fuchs:2012dt,Kong:2012ph},
\begin{equation}
\EuScript{C}_3=Z(\EuScript{C})\cong Z(\EuScript{C}_\EuScript{M}^\vee).\nn
\end{equation}
\newpage
\subsection{Frobenius algebra, full center and anyon condensation}
However, the description of the boundary theory with the module category is a little too formal. Also the relation with the ordinary 2d open-closed TQFT~\cite{Moore:2001,Segal:2001} is not clear. It would be more satisfactory if an intrinsic formulation using some structure within $\EuScript{C}$ could be established. To achieve this we need some kind of classification of the module category ${}_\EuScript{C}\EuScript{M}$. This is provided in~\cite{Ostrik:2001} through the introduction of an association algebra $A$ in $\EuScript{C}$. The category of right $A$-modules $\text{Mod}_\EuScript{C}(A)$ have a natural structure of module category over $\EuScript{C}$. In~\cite{Ostrik:2001} the inverse statement is proved: for any semisimple module category $ {}_\EuScript{C}\EuScript{M}$, there exists a semisimple algebra $A$ in $\EuScript{C}$ such that \begin{equation}
 {}_\EuScript{C}\EuScript{M}\cong \text{Mod}_\EuScript{C}(A)\nn
\end{equation}
as module categories. The algebra $A$ is determined up to Morita equivalence, and can be chosen to be connected. Connectivity means $A$ has a unique unit. Thus we could characterize the module categories with different algebras in $\EuScript{C}$.

We want to focus on the special situation when $\EuScript{C}$ is modular, to see what additional property the algebra $A$ will inherit. In this case we have (\ref{eq.C2}), which enables us to use the folding trick~\cite{Kapustin:2010if}. With this the topological boundary condition of RT-theory with $Z(\EuScript{C})$ is mapped to the topological surface operator in a RT-theory constructed directly with $\EuScript{C}$~\cite{Kapustin:2010if,Fuchs:2012dt}. The surface operator is shown manifestly, using open-string scattering diagram, to be a special symmetric Frobenius algebra~(SSFA) in the MTC $\EuScript{C}$~\cite{Kapustin:2010if}. Actually~\cite{Kapustin:2010if} gives a very nice illustration of such a complicated notion. It is then rigorously proved in~\cite{Fuchs:2012dt} that to any such surface defect there is associated a Morita equivalence class of SSFAs.

A similar analysis~\cite{Kapustin:2010if} shows that closed-string scattering diagrams define a special kind of boundary-bulk map for the given boundary condition. More precisely, it relates the boundary condition, represented by the SSFA $A$ in $\EuScript{C}$, to its derived center, which is a commutative SSFA $Z(A)$ in $Z(\EuScript{C})$. The precise definition of such a notion is provided in~\cite{Davydov:2009}, called full center of an algebra in a monoidal category. It is further proved that the full center is the property of the module category $\text{Mod}_\EuScript{C}(A)$, and thus depends only on the Morita equivalence class. Now the relation with the well-known classification of the 2d open/closed TQFT~\cite{Moore:2001,Segal:2001} becomes clear. One may say that a Morita-equivalent class of SSFAs in a MTC, together with its full center, defines an anomalous open/closed TQFT~\cite{Fuchs:2002cm}. This is the essential part of the TQFT construction of rational conformal field theory~(RCFT) in the series of papers~\cite{Fuchs:2002cm,Fuchs:2004xi,Fjelstad:2005ua}, which we denote as Fuchs-Runkel-Schweigert~(FRS)-formalism.

It will be more clear to consider the boundary-bulk map for a fixed boundary condition from the viewpoint of anyon condensation~\cite{Bais:2008ni,Bais:2008,Kong:2014qra,Hung:2014tba,Hung:2015hfa}. In particular, a systematical bootstrap analyzes of anyon condensation is performed in~\cite{Kong:2014qra} in a full categorical language. According to the analyzes, the SSFA $A$ could be considered as the vacuum of the boundary phase. The previously mentioned connectivity of $A$ guarantees that the boundary phase is stable. Boundary excitations are described by the unitary fusion category of $A$-bimodules in $\EuScript{C}$, which is again Morita equivalent to $\EuScript{C}$. The full center $Z(A)$ is the condensed set of anyons, or vacuum of the condensed phase, which is defined as a condensable algebra in $Z(\EuScript{C})$. This is an equivalent notion as a commutative SSFA in $Z(\EuScript{C})$ , with the additional physical condition that $Z(A)$ be connected. The boundary theory corresponds to the special anyon condensation from the $Z(\EuScript{C})$-phase to the trivial $\bf{1}_3$-phase. In this case, the full center/condensable algebra $Z(A)$ is a Lagrangian algebra in $Z(\EuScript{C})$, i.e.,
\begin{equation}
(\text{dim}~Z(A))^2=\text{dim}~Z(\EuScript{C}).\label{eq.Lagrangian}
\end{equation}
Here the quantum dimensions of an individual object and of the category itself are defined as usual in a MTC.

\subsection{Simplicity condition as diagonal anyon condensation}
Now we know the 4d theory could be holographically described by the boundary data of a connected SSFA $A$ in a MTC $\EuScript{C}$. But how to explicitly obtain these algebras? While the general situation could be quite complicated, there always exists a simple and special construction: the so-called Cardy case. It is also referred to as the ``charge conjugation construction''. In this case $A$ is simply taken to be the tensor unit $\bf 1$ of $\EuScript{C}$, or its bimodules in $\EuScript{C}$:
\begin{equation}
A = X\otimes X^\vee,\quad X\in \EuScript{C}.\label{eq.FA}
\end{equation}
According to~\cite{Kapustin:2010if}, these objects represent the transparent/invisible surface operators. The corresponding full center is given by
\begin{equation}
Z(A)=\oplus_{i\in I} X_i\times X_i^\vee,
\end{equation}
where $I$ labels the equivalent classes of simple objects in $\EuScript{C}$.
It is not difficult to check that the Lagrangian condition (\ref{eq.Lagrangian}) is indeed satisfied. In 2d RCFTs, this corresponds to the so-called ``diagonal modular invariant''~\cite{CFT}. Accordingly, this is called a diagonal anyon condensation for topological phases~\cite{Hung:2015hfa}.

Now let us fix $\EuScript{C}=\text{Rep}(U_{q}sl_2)$ for $q$ at roots of unity, as in the BC model. According to (\ref{eq.C2}), the bulk excitations are then given by the center $Z(\EuScript{C})\cong \EuScript{C}\boxtimes \EuScript{C}^-$, which provides a direct interpretation of the opposite braiding of the two factors in the BC model~(see also the discussion in~\cite{Yetter:98}). Moreover, the special object $A_{\text{cl}}$ (\ref{eq.simplicity}) for the simplicity condition is just $Z(A)$ in $Z(\EuScript{C})$ for the choice (\ref{eq.FA}). This explains one of the doubts I mentioned in a recent note~\cite{Zuo:2016ezr}. Similar idea has appeared recently~\cite{Han:2017geu}.

In summary, the BC model is a sub-sector of CY theory satisfying Cardy-case boundary condition on any possible $2d$ boundaries/domain walls. Alternatively speaking, the BC model is a diagonal anyon-condensed phase of the CY theory. The condensed phase is topologically trivial in the bulk, but on the $2d$ domain walls of these trivial phases, a full closed RCFT naturally emerges~\cite{KZ:1705}. A general construction of boundary/bulk RCFTs from anyon condensation is established recently~\cite{KZ:1705}. This full RCFT structure is the main topic of the next section.

\section{Stringy quantum geometry}
At the end of~\cite{Barrett:1997gw}, the authors made an interesting observation/suggestion:

"It is interesting that the sort of tensor categories that go into the state sum we are proposing
are so similar to the ones invented in constructing string theories~\cite{Moore:1988qv}. Terms in our state sum can be
interpreted as diagrams in string perturbation theory, by connecting together the diagrams we are
associating to the $4$-simplexes, and interpreting elements in the representations as string states."

In this section we want to make a first attempt towards realizing explicitly the above suggestion. Actually we have already used a lot of familiar elements from string theory or 2d conformal field theories~(CFTs): open-string/closed-string scattering diagrams, diagonal modular invariants, et al. And many mathematical notions we have employed, such as module category, associative algebra in a monoidal category, and full center, are motivated (or partially motivated) by the TQFT construction of RCFTs~\cite{Fuchs:2002cm}. Thus one would expect that the correspondence between general types of anyon condensation and modular invariants should somehow be a consequence, rather than a coincidence~\cite{Bais:2008ni,Hung:2015hfa}. Indeed, the exact relation has been established recently in the general case~\cite{KZ:1705}. Here we only focus on the special case relevant for the BC model.

Up to now we have been focusing on the MTC $\EuScript{C}=\text{Rep}(U_{q}sl_2)$ for $q$ at roots of unity. For many years people believe that such a MTC, for the representations of the quantum deformation of a general simple Lie algebra $g$, is equivalent to the MTC $\mathcal{C}_{\hat g}$ of modules of the affine Lie algebra $\hat g$ at level $k$. However, it takes quite a long time for the mathematicians to rigorously prove this. Only until recently it becomes clear that the complete proof for the general situation could indeed be achieved~\cite{Huang:2013jza}. Let us recall the precise statement as follows~\cite{Kazhdan-Lusztig:I,Kazhdan-Lusztig:II,Kazhdan-Lusztig:III,Kazhdan-Lusztig:IV,Finkelberg:1996,Finkelberg:2013}
\cite{Huang:2005,Huang:2005gs,Huang:2013jza}:

For a general simple Lie algebra $g$, the following equivalence
\begin{equation}
\text{Rep}(U_q g)\cong \mathcal{C}_{{\hat g}},
\end{equation}
as MTCs exists, for
\begin{equation}
q=\mathe^{\mathi \pi/m(k+h^\vee)}.
\end{equation}
\vspace{2.5mm}
Here $m$ is squared ratio of a long root and a short root, and $h^\vee$ is dual Coxeter number. See~\cite{mathoverflow:177519,mathoverflow:178113} for some interesting discussions on this theorem. As emphasized in~\cite{Huang:2013jza}, the construction of the tensor category structure for the modules of the affine Lie algebra is indeed a vertex-algebraic problem, which requires the general theory of Vertex Operator Algebra~(VOA). While on the quantum group side, it is still a pure algebraic construction based on Lie algebra representation theory. That may explain why it takes such a long time to finally establish the MTC structure of the modules of affine Lie algebras for general simple Lie algebras.

Now with the above theorem, we could consider all the data in constructing the BC model as from the category of the modules of the affine Lie algebra $\widehat{\mathfrak{su}}(2)_k$ with positive integer level $k$~\footnote[5]{When we consider open $(1+1)d$ boundary supporting chiral gapless excitations, then this enrichment would be requirable rather than desirable~\cite{KZ:1705}.}. As the loop extension of the Lie algebra, affine Lie algebra allows us to include a rich local structure into the state sum, in contrast with the quantum group. In fact, the affine Lie algebra $\widehat{\mathfrak{su}}(2)_k$ together with a Morita-equivalence class of the SSFA (\ref{eq.FA}) allows us to construct a full CFT, namely Wess-Zumino-Witten model, on any surfaces~\cite{Fuchs:2002cm,Fjelstad:2005ua}. The advantage of such a construction is that it is expected to be completely background independent/coordinate free. In~\cite{Kong:2011jf}, it is proposed that a full 2d CFT could be viewed as a ``stringy algebraic geometry''. Here we could imitate this notion and call the 4d spacetime with full CFT structure on the 2d surfaces as a ``stringy quantum geometry''. In the following I will try to sketch some attractive features of this new geometry.

We already see that the BC model is completely fixed once we choose for all the faces, original or intermediate, the special object (\ref{eq.simplicity}). This is because the $U_{q}sl_2$ intertwiner for three simple objects is fixed once the normalization for all the simple objects are chosen~\cite{Reisenberger:1998bn}. While this looks very elegant, it also becomes a fatal weakness of the model: it lacks degrees of freedom to accommodate some allowed quantum data like the face angles~\cite{Baez:1999tk,Engle:2007uq}. Also it is suspected that the correlations between neighboring simplices are too limited~\cite{Baez:1999tk,Freidel:2007py}. Various relaxations to the original model have been made to improve these situation~\cite{Engle:2007uq,Freidel:2007py,Engle:2007wy}. Here the introduction of the CFT structure may provide a new solution, and without modifying the elegant topological structure.

First at the topological level, we have just reinterpreted the original model with some new mathematical structure, but with no essential changes. In particular, the whole structure is still fully constrained by the three-valent intertwiners, or structure constants of the three-point correlation functions in RCFT. This also has a nice interpretation in the new mathematical setting: they are given by the fusing matrices of the category of the $A$-bimodules~\cite{Fuchs:2004xi}. For the Cardy case, the explicit expressions for both the fusing matrices and the structure constants are derived~\cite{Fuchs:2004xi}.

But the affine Lie algebra, viewed as a VOA, has in addition to the topological structure also complex analytic structure. These may enable us to capture the required local correlations and quantum data. First consider the faces, or the dual links. At the tensor category level, they are colored by (\ref{eq.simplicity}) as the direct sum of simple objects. But each simple object corresponds to a conformal family, which possesses an infinite tower of descendants. For a large excitation level, the number of descendants increases exponentially with the level. Based on this, it is suggested in~\cite{Zuo:2016ezr} that this huge correlation across the faces could explain the Bekenstein-Hawking formula for the vacuum entanglement entropy. These descendants are completely hidden in the two and three-point conformal blocks. In other words, the corresponding two- and three-point functions have universal dependence on the coordinates of insertion points, which are completely fixed by conformal invariance. The situation changes for the four-point conformal blocks, in which the contributions of different descendants could be distinguished by the different coordinate dependence. This signifies that the tetrahedra, or more properly four-punctured sphere, has its own degrees of freedom not fixed by the structure constants. Further gluing five such punctured spheres, one obtains a fattened $4$-simplex. This is a genus-$6$ Riemannian surface, as shown in~\cite{Haggard:2014xoa,Haggard:2015yda,Haggard:2015kew}. Correspondingly, on the CFT side one describes this with the vacuum partition function on this surface. One would expect the dependence of the partition function on the moduli space captures the underlying geometry. One could continue to glue different fattened $4$-cells across punctured spheres. It is a little difficult to imagine the process, but a very nice picture is provided in the FRS-formalism~\cite{Fuchs:2002cm}. One makes the double cover of the punctured sphere, with the punctures identified. Then one could easily connect the two pieces to the two neighboring fattened $4$-cells. Since the punctured spheres have their own degrees of freedom, one expects that these data would be transmitted during this gluing procedure. In summary, in the FRS-formalism we have very nice factorization/sewing properties for the CFT correalators, which perfectly match this gluing procedure for the spacetime geometry.

It would be interesting to notice that the above situation is quite similar to the recent Witten-Costello construction of integrable lattice model~\cite{Costello:2013sla,Witten:2016spx}. Starting from the CS theory, they introduce the loop extension (without central extension) of the gauge group. By properly recombining the loop parameter and one space coordinate, they obtain a mixed holomorphic/topological theory, which results the long-expected full-fledged Yang-Baxter equation with the spectral parameter~\cite{Yang:1967bm,Baxter:1972hz}. Since the loop parameter mixes with one of the space coordinate, the original three dimensional symmetry in CS theory is reduced to the 2-dimensional symmetry in the integrable lattice models~\cite{Costello:2013sla,Witten:2016spx}.

The comparison with the Witten-Costello construction would be quite inspiring for the determination of the symmetry algebra here.  Eventually this provides an alternative interpretation of the decomposition (\ref{eq.Spin}). So I would like to discuss a little more of the choice of the Lie algebra at this point, which is unfortunately ignored in the first version of~\cite{Zuo:2016ezr}. More generally, this would be related to the notion of so-called quantum symmetry. While the original internal symmetry group is $\text{SO}(4)$ in 4d Euclidean case, after loop and central extension we are left with $\hat g =\widehat{\mathfrak{su}}(2)_k$, with $\mathfrak{su}(2)$ the symmetry algebra and $\widehat{\mathfrak{su}}(2)_k$ the spectrum-generating algebra. To get a complete theory with the symmetry, we need to consider the full field algebra $\hat g\otimes \hat g$~\cite{Huang:2005gz}.
Then the corresponding modules
\begin{equation}
\mathcal{C}_{\hat g \otimes \hat g} \cong \mathcal{C}_{\hat g} \boxtimes (\mathcal{C}_{\hat g})^- \cong Z(\mathcal{C}_{\hat g})\label{eq.CV2}
\end{equation}
provide the required bulk excitations. Similar consideration for the group structure of the BC model has been noticed in the early literature~\cite{DePietri:1999bx}.

If one accepts this logic, then it will be natural that in the Lorentzian case the Lie algebra $\mathfrak{su}(1,1)$ is to be used. See \cite{Carlip:2014bfa,Liu:2017bfk} for recent discuss on such a choice. Extending to the affine Lie algebra and making the doubling to the full field algebra then gives $\widehat{\mathfrak{su}}(1,1)_k\otimes\widehat{\mathfrak{su}}(1,1)_k$. This is the starting point of~\cite{Zuo:2016ezr}, which is chosen there simply by analogy with the investigation of $(2+1)d$ black holes~\cite{Carlip:1994gy}. However, compared to the Euclidean case there is a crucial difference here. Now $\widehat{\mathfrak{su}}(1,1)_k$ is not rational any more, and correspondingly the category of modules do not have the structure of a MTC. Then the relations (\ref{eq.C2}) and (\ref{eq.CV2}) do not hold. The folding trick and the surface operator/boundary condition correspondence also become ambiguous. But from the viewpoint of the 2d CFT, this may still be the natural choice. Also the transparent objects (\ref{eq.FA}) and the corresponding diagonal modular invariant may still exist. If this is accepted, one could then follow the derivation there to get the Bekenstein-Hawking formula for the vacuum entanglement entropy. It should be pointed out that no simplicity condition has been imposed in such a derivation (which is of course not a proper choice)~\cite{Zuo:2016ezr}. But it is easy to show that further imposing this changes at most the sub-leading logarithmic term.

An immediate question will be, when will the 4d rotational/Lorentzian symmetry be restored? And a similar question could be asked in the Witten-Costello construction, in which part of answer is already known. We leave this question to the future study.

\section{Discussion}
In this paper I have illustrated that the simplicity condition in the BC model corresponds to the special symmetric Frobenius algebra in $\EuScript{C}=\text{Rep}(U_{q}sl_2)$ for the Cardy case. In other words, it corresponds to a diagonal anyon condensation from the original CY topological phase characterized by $Z(\EuScript{C})\cong \EuScript{C} \boxtimes \EuScript{C}^-$. The equivalence for the MTC structure between quantum groups and the affine Lie algebras allows us to reinterpret the whole theory with the modules of the affine Lie algebra $\widehat{\mathfrak{su}}(2)_k$. Employing the general VOA framework and the FRS-formalism, this enables us to construct a full CFT on any surfaces. Essentially this makes it possible to introduce local geometrical data into the spin-foam framework, which has been one of the long-awaited goal of quantum gravity~\cite{Baez:1999sr}. I have roughly sketched the geometrical structure with the usual simplicial decomposition, which is also the strategy adopted in the FRS formalism. As I mentioned, perhaps the handle-decomposition could provide a better and more general description of the geometry, just as of the topological structure~\cite{Roberts:1995}. The topological theory based on handle-decomposition has been recently developed in~\cite{Barenz:2016nzn}. It would be interesting to further enrich the chain-mail diagram in this description with VOA data.

We have given some arguments suggesting that in the Lorentz case we should follow the same procedure with $\widehat{\mathfrak{su}}(2)_k$ replaced by $\widehat{\mathfrak{su}}(1,1)_k$. Unfortunately, the affine Lie algebra is no longer rational any more. So many nice properties of those mathematical structure in the rational case are lost. One even does not know if the whole scheme is still feasible. Some preliminary analysis has been made in my recent note~\cite{Zuo:2016ezr}, as I mentioned. This is possible due to the observation that, in the large-$k$ limit, the resulting theory could be formally viewed as a gauge-fixed version of $(2+1)d$ closed-string theory. Some familiar techniques from string theory are then employed to make the state-counting for the surface degrees of freedom. It is hoped that by fusing together the combinatoric/topological construction with the geometric aspect of string theory, further progress could be made. As we show, this is also what Barrett and Crane have suggested/expected almost 20 years ago.

\section*{Acknowledgments}
Much of this work has been done during my visit at Yau Mathematical Sciences Center, Tsinghua University. The hospitality of YMSC is acknowledged. I want to thank Si Li in particular for the invitation, for explaining many relevant mathematical notions, and for encouraging me to complete the subject. I am grateful to Liang Kong and Hao Zheng for sending me the updated version of their recent work and illustrating some new results therein. I have also benefitted from the discussions with Yi-Hong Gao, Ling-Yan Hung, Wei Song, Jun-Bao Wu, Jian Zhou and many others. The work was partially supported by the National Natural Science Foundation of China under Grant No. 11405065.


\begin{thebibliography}{10}

\bibitem{Segal:1987sk}
G.~B. Segal.
\newblock {\em {THE DEFINITION OF CONFORMAL FIELD THEORY}}.
\newblock In {\em {IN *COMO 1987, PROCEEDINGS, DIFFERENTIAL GEOMETRICAL METHODS
  IN THEORETICAL PHYSICS* 165-171.}}, 1987.

\bibitem{Atiyah:1989vu}
M.~Atiyah, {\it TOPOLOGICAL QUANTUM FIELD THEORIES},
\newblock Inst. Hautes Etudes Sci. Publ. Math. \textbf{68} (1989): 175-186.

\bibitem{Baez:1995xq}
J.~C. Baez and J.~Dolan, {\it Higher-dimensional algebra and topological quantum field theory},
\newblock J. Math. Phys. \textbf{36} (1995): 6073-6105
  [\href{http://arxiv.org/abs/q-alg/9503002}{arXiv: q-alg/9503002}].

\bibitem{Baez:1995ph}
John~C. Baez, {\it Four-Dimensional $BF$ Theory as a Topological
Quantum Field Theory},
\newblock Lett. Math. Phys. \textbf{38} (1996): 129-143
  [\href{http://arxiv.org/abs/q-alg/9507006}{arXiv: q-alg/9507006}].

\bibitem{Baez:1999sr}
J.~C. Baez,{\it An Introduction to Spin Foam Models
of $BF$ Theory and Quantum Gravity},
\newblock Lect. Notes Phys. \textbf{543} (2000): 25-94
  [\href{http://arxiv.org/abs/gr-qc/9905087}{arXiv: gr-qc/9905087}].

\bibitem{Ashtekar:1986yd}
A.~Ashtekar, {\it New Variables for Classical and Quantum Gravity},
\newblock Phys. Rev. Lett. \textbf{57} (1986): 2244-2247.

\bibitem{Barrett:1997gw}
John~W. Barrett and Louis Crane, {\it Relativistic spin networks and quantum gravity},
\newblock J. Math. Phys. \textbf{39} (1998): 3296-3302
  [\href{http://arxiv.org/abs/gr-qc/9709028}{arXiv: gr-qc/9709028}].

\bibitem{Crane:1993if}
Louis Crane and David Yetter, {\it A CATEGORICAL CONSTRUCTION OF 4D
TOPOLOGICAL QUANTUM FIELD THEORIES},
\newblock {In {\it *Dayton 1992, Proceedings, Quantum topology*} 120-130}
  (1993): [\href{http://arxiv.org/abs/hep-th/9301062}{arXiv: hep-th/9301062}].

\bibitem{Ponzano-Regge}
G.~Ponzano and T.~Regge.
\newblock {\it Spectroscopic and Group Theoretical Methods in Physics}, edited
  by F. Bloch ~(North-Holland, Amsterdam, 1968), pp. 1--58.

\bibitem{Turaev:1992hq}
V.~G. Turaev and O.~Y. Viro, {\it State sum invariants of $3$-manifolds and
quantum $6j$-symbols},
\newblock Topology \textbf{31} (1992): 865-902.

\bibitem{Turaev:1992}
V.~G. {Turaev}, {\it Modular categories and 3-manifold invariants},
\newblock International Journal of Modern Physics B \textbf{6} (1992):
  1807-1824.

\bibitem{Yetter:98}
D.~N. {Yetter}, {\it Generalized Barrett-Crane Vertices and Invariants of Embedded Graphs},
\newblock ArXiv Mathematics e-prints (1998):
  [\href{http://arxiv.org/abs/math/9801131}{arXiv: math/9801131}].

\bibitem{Reisenberger:1998bn}
Michael~P. Reisenberger, {\it On relativistic spin network vertices
}, \newblock J. Math. Phys. \textbf{40} (1999): 2046-2054
  [\href{http://arxiv.org/abs/gr-qc/9809067}{arXiv: gr-qc/9809067}].

\bibitem{Reisenberger:1998fk}
Michael~P. Reisenberger, {\it Classical Euclidean general relativity from ``left-handed area = right-handed area''},
\newblock Class. Quant. Grav. \textbf{16} (1999): 1357
  [\href{http://arxiv.org/abs/gr-qc/9804061}{arXiv: gr-qc/9804061}].

\bibitem{Roberts:1995}
Justin Roberts, {\it SKEIN THEORY AND TURAEV-VIRO INVARIANTS},
\newblock Topology \textbf{34} (1995): 771 - 787.

\bibitem{Lurie:2009}
J.~{Lurie}, {\it On the Classification of Topological Field Theories
}, \newblock ArXiv e-prints (2009): [\href{http://arxiv.org/abs/0905.0465}{arXiv:
  0905.0465}].

\bibitem{nLab-holography}
{\it Holographic principle of higher category theory},
  https://ncatlab.org/nlab/show/holographic+principle+of+higher+category+theory.

\bibitem{Kitaev:1997wr}
A.~{\relax Yu}. Kitaev, {\it Fault-tolerant quantum computation by anyons
}, \newblock Annals Phys. \textbf{303} (2003): 2-30
  [\href{http://arxiv.org/abs/quant-ph/9707021}{arXiv: quant-ph/9707021}].

\bibitem{Levin:2004mi}
Michael~A. Levin and Xiao-Gang Wen, {\it String-net condensation: A physical mechanism for topological phases},
\newblock Phys. Rev. \textbf{B71} (2005): 045110
  [\href{http://arxiv.org/abs/cond-mat/0404617}{arXiv: cond-mat/0404617}].

\bibitem{Walker-Wang:2011}
K.~{Walker} and Z.~{Wang}, {\it (3+1)-TQFTs and Topological Insulators
}, \newblock Frontiers of Physics \textbf{7} (2012): 150-159
  [\href{http://arxiv.org/abs/1104.2632}{arXiv: 1104.2632}].

\bibitem{Kong:2014qka}
Liang Kong and Xiao-Gang Wen, {\it Braided fusion categories, gravitational anomalies, and the mathematical framework for topological orders in any dimensions},
\newblock ArXiv e-prints (2014):
  [\href{http://arxiv.org/abs/1405.5858}{arXiv: 1405.5858}].

\bibitem{Kong:2015}
L.~{Kong}, X.-G. {Wen}, and H.~{Zheng}, {\it Boundary-bulk relation for topological orders as the functor mapping higher categories to their centers}, \newblock ArXiv e-prints (2015): [\href{http://arxiv.org/abs/1502.01690}{arXiv: 1502.01690}].

\bibitem{Kong:2017}
L.~{Kong}, X.-G. {Wen}, and H.~{Zheng}, {\it Boundary-bulk relation in topological orders
}, \newblock ArXiv e-prints (2017): [\href{http://arxiv.org/abs/1702.00673}{arXiv: 1702.00673}].

\bibitem{Witten:1988hf}
Edward Witten, {\it Quantum Field Theory and the Jones Polynomial},
\newblock Commun. Math. Phys. \textbf{121} (1989): 351-399.

\bibitem{Reshetikhin:1991}
N.~Reshetikhin and V.~G. Turaev, {\it Invariants of $3$-manifolds via link polynomials
and quantum groups},
\newblock Inventiones mathematicae \textbf{103} (1991): 547--597.

\bibitem{Barrett:2004im}
John~W. Barrett, J.~Manuel Garcia-Islas, and Joao~Faria Martins, {\it Observables in the Turaev-Viro and Crane-Yetter models
}, \newblock J. Math. Phys. \textbf{48} (2007): 093508
  [\href{http://arxiv.org/abs/math/0411281}{arXiv: math/0411281}].

\bibitem{KZ:1705}
L.~{Kong} and H.~{Zheng}, {\it Gapless edges of 2d topological orders and enriched monoidal categories
}, \newblock ArXiv e-prints (2017): [\href{http://arxiv.org/abs/1705.01087}{arXiv:
  1705.01087}].

\bibitem{KZ:1704}
L.~{Kong} and H.~{Zheng}, {\it Drinfeld center of enriched monoidal categories}, \newblock ArXiv e-prints (2017): [\href{http://arxiv.org/abs/1704.01447}{arXiv:
  1704.01447}].

\bibitem{Freed:2012}
D.~S. Freed, {\it 4-3-2-8-7-6},
\newblock talk at {\it ASPECTS of Topology} (2012):.

\bibitem{Crane:1994ji}
Louis Crane, Louis~H. Kauffman, and David~N. Yetter, {\it State-Sum Invariants of 4-Manifolds I},
\newblock {Journal of Knot Theory and its Ramifications} 6.2, (1997): pp.
  177--234. [\href{http://arxiv.org/abs/hep-th/9409167}{arXiv:
  hep-th/9409167}].

\bibitem{Williamson:2016evv}
Dominic~J. Williamson and Zhenghan Wang, {\it Hamiltonian models for topological phases of matter in three spatial dimensions},
\newblock Annals of Physics \textbf{377} (2017): 311 - 344
  [\href{http://arxiv.org/abs/1606.07144}{arXiv: 1606.07144}].

\bibitem{Crane:1993cm}

\newblock Louis Crane, Louis~H. Kauffman, and David Yetter, {\it Evaluating the Crane-Yetter Invariant},
 \newblock ArXiv e-prints (1993):
  [\href{http://arxiv.org/abs/hep-th/9309063}{arXiv: hep-th/9309063}].

\bibitem{Smolin:1995vq}
Lee Smolin, {\it Linking Topological Quantum Field Theory and Nonperturbative Quantum Gravity},
\newblock J. Math. Phys. \textbf{36} (1995): 6417-6455
  [\href{http://arxiv.org/abs/gr-qc/9505028}{arXiv: gr-qc/9505028}].

\bibitem{Kirillov:2010nh}

\newblock Alexander Kirillov, Jr. and Benjamin Balsam, {\it Turaev-Viro invariants as an extended TQFT},  
\newblock ArXiv e-prints (2010):
  [\href{http://arxiv.org/abs/1004.1533}{arXiv: 1004.1533}].

\bibitem{Turaev:2010}
V.~{Turaev} and A.~{Virelizier}, {\it On two approaches to 3-dimensional TQFTs},
\newblock ArXiv e-prints (2010): [\href{http://arxiv.org/abs/1006.3501}{arXiv:
  1006.3501}].

\bibitem{Balsam:2010}
B.~{Balsam}, {\it Turaev-Viro invariants as an extended TQFT II},
\newblock ArXiv e-prints (2010): [\href{http://arxiv.org/abs/1010.1222}{arXiv:
  1010.1222}].

\bibitem{Balsam:2010B}
B.~{Balsam}, {\it Turaev-Viro invariants as an extended TQFT III},
\newblock ArXiv e-prints (2010): [\href{http://arxiv.org/abs/1012.0560}{arXiv:
  1012.0560}].

\bibitem{Fuchs:2012dt}
Jurgen Fuchs, Christoph Schweigert, and Alessandro Valentino, {\it Bicategories for boundary conditions and for surface defects in 3-d TFT}, \newblock Commun. Math. Phys. \textbf{321} (2013): 543-575
  [\href{http://arxiv.org/abs/1203.4568}{arXiv: 1203.4568}].

\bibitem{Joyal:1991}
Andr{\'e} Joyal and Ross Street.
\newblock {\em An introduction to Tannaka duality and quantum groups},
  413--492.
\newblock Springer Berlin Heidelberg, Berlin, Heidelberg, 1991.

\bibitem{Buerschaper:2009}
O.~{Buerschaper} and M.~{Aguado}, {\it Mapping Kitaev's quantum double lattice models to Levin and Wen's string-net models},
\newblock \prb \textbf{80} (2009): 155136
  [\href{http://arxiv.org/abs/0907.2670}{arXiv: 0907.2670}].

\bibitem{Kadar:2009fs}
Zoltan Kadar, Annalisa Marzuoli, and Mario Rasetti, {\it Microscopic description of 2d topological phases, duality and 3d state sums},
\newblock Adv. Math. Phys. \textbf{2010} (2010): 671039
  [\href{http://arxiv.org/abs/0907.3724}{arXiv: 0907.3724}].

\bibitem{Buerschaper:2010yf}
Oliver Buerschaper, Matthias Christandl, Liang Kong, and Miguel Aguado, {\it Electric-magnetic duality of lattice systems with topological order},
\newblock Nucl. Phys. \textbf{B876} (2013): 619-636
  [\href{http://arxiv.org/abs/1006.5823}{arXiv: 1006.5823}].

\bibitem{Kirillov:2011}
A.~{Kirillov}, Jr, {\it String-net model of Turaev-Viro invariants},
\newblock ArXiv e-prints (2011): [\href{http://arxiv.org/abs/1106.6033}{arXiv:
  1106.6033}].

\bibitem{Kitaev-Kong:2012}
A.~{Kitaev} and L.~{Kong}, {\it Models for gapped boundaries and domain walls},
 \newblock Communications in Mathematical Physics \textbf{313} (2012): 351-373
  [\href{http://arxiv.org/abs/1104.5047}{arXiv: 1104.5047}].

\bibitem{Kong:2012ph}
Liang Kong, {\it Some universal properties of Levin-Wen models},
\newblock Proceedings of XVIITH International Congress of Mathematical Physics,
  444-455 (2014): [\href{http://arxiv.org/abs/1211.4644}{arXiv: 1211.4644}].

\bibitem{Ostrik:2001}
V.~{Ostrik}, {\it Module categories, weak Hopf algebras and modular invariants},
\newblock Transformation Groups \textbf{8} (2003): 177--206
  [\href{http://arxiv.org/abs/math/0111139}{arXiv: math/0111139}].

\bibitem{Moore:2001}
Gregory Moore, {\it Lectures on Branes, K-theory and RR Charges},
\newblock {Lecture notes from the Clay Institute School on Geometry and
  String Theory held at the Isaac Newton Institute}, Cambridge, UK.
  (2001--2002):.

\bibitem{Segal:2001}
G.~Segal, {\it Topological structures in string theory},
\newblock Philosophical Transactions of the Royal Society of London A:
  Mathematical, Physical and Engineering Sciences \textbf{359} (2001):
  1389--1398.

\bibitem{Kapustin:2010if}
Anton Kapustin and Natalia Saulina, {\it Surface operators in 3d Topological Field Theory and 2d Rational Conformal Field Theory},
\newblock in Hisham Sati, Urs Schreiber (eds.), Mathematical Foundations
  of Quantum Field and Perturbative String Theory, Proceedings in Symposia in
  Pure Mathematics, volume 83 AMS (2011):
  [\href{http://arxiv.org/abs/1012.0911}{arXiv: 1012.0911}].

\bibitem{Davydov:2009}
A.~{Davydov}, {\it Centre of an algebra},
\newblock Advances in Mathematics \textbf{225} (2010): 319 - 348
  [\href{http://arxiv.org/abs/0908.1250}{arXiv: 0908.1250}].

\bibitem{Fuchs:2002cm}
Jurgen Fuchs, Ingo Runkel, and Christoph Schweigert, {\it TFT construction of RCFT correlators I: Partition functions},
\newblock Nucl. Phys. \textbf{B646} (2002): 353-497
  [\href{http://arxiv.org/abs/hep-th/0204148}{arXiv: hep-th/0204148}].

\bibitem{Fuchs:2004xi}
Jurgen Fuchs, Ingo Runkel, and Christoph Schweigert, {\it TFT construction of RCFT correlators IV: Structure constants and correlation functions},
\newblock Nucl. Phys. \textbf{B715} (2005): 539-638
  [\href{http://arxiv.org/abs/hep-th/0412290}{arXiv: hep-th/0412290}].

\bibitem{Fjelstad:2005ua}
Jens Fjelstad, Jurgen Fuchs, Ingo Runkel, and Christoph Schweigert, {\it TFT construction of RCFT correlators V: Proof of modular invariance and factorisation},
\newblock Theor. Appl. Categor. \textbf{16} (2006): 342-433
  [\href{http://arxiv.org/abs/hep-th/0503194}{arXiv: hep-th/0503194}].

\bibitem{Bais:2008ni}
F.~A. Bais and J.~K. Slingerland, {\it Condensate induced transitions between topologically ordered phases},
\newblock Phys. Rev. \textbf{B79} (2009): 045316
  [\href{http://arxiv.org/abs/0808.0627}{arXiv: 0808.0627}].

\bibitem{Bais:2008}
F.~A. {Bais}, J.~K. {Slingerland}, and S.~M. {Haaker}, {\it A theory of topological edges and domain walls},
\newblock Physical Review Letters \textbf{102} (2009): 220403
  [\href{http://arxiv.org/abs/0812.4596}{arXiv: 0812.4596}].

\bibitem{Kong:2014qra}
Liang Kong, {\it Anyon condensation and tensor categories},
\newblock Nucl. Phys. \textbf{B886} (2014): 436-482
  [\href{http://arxiv.org/abs/1307.8244}{arXiv: 1307.8244}].

\bibitem{Hung:2014tba}
Ling-Yan Hung and Yidun Wan, {\it Ground State Degeneracy of Topological Phases on Open Surfaces},
\newblock Phys. Rev. Lett. \textbf{114} (2015): 076401
  [\href{http://arxiv.org/abs/1408.0014}{arXiv: 1408.0014}].

\bibitem{Hung:2015hfa}
Ling-Yan Hung and Yidun Wan, {\it Generalized ADE Classification of Gapped Domain Walls},
\newblock JHEP \textbf{07} (2015): 120
  [\href{http://arxiv.org/abs/1502.02026}{arXiv: 1502.02026}].

\bibitem{CFT}
Philippe Di~Francesco, Pierre Mathieu, and David Senechal.
\newblock {\em Conformal Field Theory}.
\newblock Springer, 1997.

\bibitem{Zuo:2016ezr}
Fen Zuo, {\it A note on the architecture of spacetime geometry},
\newblock ArXiv e-prints (2016): [\href{http://arxiv.org/abs/1607.05866}{arXiv:
  1607.05866}].

\bibitem{Han:2017geu}
Muxin Han and Zichang Huang, {\it Loop-Quantum-Gravity Simplicity Constraint as Surface Defect in Complex Chern-Simons Theory},
\newblock Phys. Rev. \textbf{D95} (2017): 104031
  [\href{http://arxiv.org/abs/1702.03285}{arXiv: 1702.03285}].

\bibitem{Moore:1988qv}
Gregory~W. Moore and Nathan Seiberg, {\it Classical and Quantum Conformal Field Theory},
\newblock Commun. Math. Phys. \textbf{123} (1989): 177.

\bibitem{Huang:2013jza}
Yi-Zhi Huang and James Lepowsky, {\it Tensor categories and the mathematics of rational and logarithmic conformal field theory},
\newblock J. Phys. \textbf{A46} (2013): 494009
  [\href{http://arxiv.org/abs/1304.7556}{arXiv: 1304.7556}].

\bibitem{Kazhdan-Lusztig:I}
D.~Kazhdan and G.~Lusztig, {\it TENSOR STRUCTURES ARISING FROM AFFINE LIE ALGEBRAS. I},
\newblock J. Amer. Math. Soc. \textbf{6} (1993): 905--947.

\bibitem{Kazhdan-Lusztig:II}
D.~Kazhdan and G.~Lusztig, {\it TENSOR STRUCTURES ARISING FROM AFFINE LIE ALGEBRAS. II},
\newblock J. Amer. Math. Soc. \textbf{6} (1993): 949--1011.

\bibitem{Kazhdan-Lusztig:III}
D.~Kazhdan and G.~Lusztig, {\it TENSOR STRUCTURES ARISING FROM AFFINE LIE ALGEBRAS. III},
\newblock J. Amer. Math. Soc. \textbf{7} (1994): 335--381.

\bibitem{Kazhdan-Lusztig:IV}
D.~Kazhdan and G.~Lusztig, {\it TENSOR STRUCTURES ARISING FROM AFFINE LIE ALGEBRAS. IV},
\newblock J. Amer. Math. Soc. \textbf{7} (1994): 383--453.

\bibitem{Finkelberg:1996}
M.~Finkelberg, {\it AN EQUIVALENCE OF FUSION CATEGORIES},
\newblock Geom. Funct. Anal. \textbf{6} (1996): 249--267.

\bibitem{Finkelberg:2013}
M.~Finkelberg, {\it ERRATUM TO: AN EQUIVALENCE OF FUSION CATEGORIES},
\newblock Geom. Funct. Anal. \textbf{23} (2013): 810--811.

\bibitem{Huang:2005}
Yi-Zhi Huang, {\it Vertex operator algebras, the Verlinde conjecture,
and modular tensor categories},
\newblock Proceedings of the National Academy of Sciences of the United States
  of America \textbf{102} (2005): 5352-5356
  [\href{http://arxiv.org/abs/http://www.pnas.org/content/102/15/5352.full.pdf}{arXiv:
  http://www.pnas.org/content/102/15/5352.full.pdf}].

\bibitem{Huang:2005gs}
Yi-Zhi Huang, {\it Rigidity and modularity of vertex tensor categories},
\newblock Commun. Contemp. Math. \textbf{10} (2008): 871--911
  [\href{http://arxiv.org/abs/math/0502533}{arXiv: math/0502533}].

\bibitem{mathoverflow:177519}
https://mathoverflow.net/questions/177519/why-should-affine-lie-algebras-and-quantum-groups-have-equivalent-representation.

\bibitem{mathoverflow:178113}
https://mathoverflow.net/questions/178113/whats-the-state-of-affairs-concerning-the-identification-between-quantum-group.

\bibitem{Kong:2011jf}
Liang Kong, {\it Conformal field theory and a new geometry},
\newblock in Hisham Sati, Urs Schreiber (eds.), Mathematical Foundations
  of Quantum Field and Perturbative String Theory, Proceedings in Symposia in
  Pure Mathematics, volume 83 AMS (2011):
  [\href{http://arxiv.org/abs/1107.3649}{arXiv: 1107.3649}].

\bibitem{Baez:1999tk}
John~C. Baez and John~W. Barrett, {\it The Quantum Tetrahedron in 3 and 4 Dimensions},
\newblock Adv. Theor. Math. Phys. \textbf{3} (1999): 815-850
  [\href{http://arxiv.org/abs/gr-qc/9903060}{arXiv: gr-qc/9903060}].

\bibitem{Engle:2007uq}
Jonathan Engle, Roberto Pereira, and Carlo Rovelli, {\it The loop-quantum-gravity vertex-amplitude},
\newblock Phys. Rev. Lett. \textbf{99} (2007): 161301
  [\href{http://arxiv.org/abs/0705.2388}{arXiv: 0705.2388}].

\bibitem{Freidel:2007py}
Laurent Freidel and Kirill Krasnov, {\it A New Spin Foam Model for 4d Gravity},
\newblock Class. Quant. Grav. \textbf{25} (2008): 125018
  [\href{http://arxiv.org/abs/0708.1595}{arXiv: 0708.1595}].

\bibitem{Engle:2007wy}
Jonathan Engle, Etera Livine, Roberto Pereira, and Carlo Rovelli, {\it LQG vertex with finite Immirzi parameter},
\newblock Nucl. Phys. \textbf{B799} (2008): 136-149
  [\href{http://arxiv.org/abs/0711.0146}{arXiv: 0711.0146}].

\bibitem{Haggard:2014xoa}
Hal~M. Haggard, Muxin Han, Wojciech Kami\'{n}ski, and Aldo Riello, {\it SL(2,C) Chern-Simons theory, a non-planar graph operator, and 4D quantum gravity with a cosmological constant: Semiclassical geometry},
\newblock Nucl. Phys. \textbf{B900} (2015): 1-79
  [\href{http://arxiv.org/abs/1412.7546}{arXiv: 1412.7546}].

\bibitem{Haggard:2015yda}
Hal~M. Haggard, Muxin Han, Wojciech Kami\'{n}ski, and Aldo Riello, {\it Four-dimensional Quantum Gravity with a Cosmological Constant from Three-dimensional Holomorphic Blocks},
\newblock Phys. Lett. \textbf{B752} (2016): 258-262
  [\href{http://arxiv.org/abs/1509.00458}{arXiv: 1509.00458}].

\bibitem{Haggard:2015kew}

Hal~M. Haggard, Muxin Han, Wojciech Kami\'{n}ski, and Aldo Riello, {\it SL(2,C) Chern-Simons Theory, Flat Connections, and Four-dimensional Quantum Geometry},
\newblock  ArXiv e-prints (2015): [\href{http://arxiv.org/abs/1512.07690}{arXiv: 1512.07690}].

\bibitem{Costello:2013sla}
Kevin Costello, {\it Integrable lattice models from four-dimensional field theories},
\newblock Proc. Symp. Pure Math. \textbf{88} (2014): 3-24
  [\href{http://arxiv.org/abs/1308.0370}{arXiv: 1308.0370}].

\bibitem{Witten:2016spx}
Edward Witten, {\it Integrable Lattice Models From Gauge Theory},
\newblock  ArXiv e-prints (2016): [\href{http://arxiv.org/abs/1611.00592}{arXiv:
  1611.00592}].

\bibitem{Yang:1967bm}
Chen-Ning Yang, {\it Some Exact Results for the Many-Body Problem in one Dimension with Repulsive Delta-Function Interaction},
\newblock Phys. Rev. Lett. \textbf{19} (1967): 1312-1314.

\bibitem{Baxter:1972hz}
Rodney~J. Baxter, {\it Partition function of the Eight-Vertex lattice model},
\newblock Annals Phys. \textbf{70} (1972): 193-228.
\newblock [Annals Phys.281,187(2000)].

\bibitem{Huang:2005gz}
Yi-Zhi Huang and Liang Kong, {\it Full field algebras},
\newblock Commun. Math. Phys. \textbf{272} (2007): 345-396
  [\href{http://arxiv.org/abs/math/0511328}{arXiv: math/0511328}].

\bibitem{DePietri:1999bx}
Roberto De~Pietri, Laurent Freidel, Kirill Krasnov, and Carlo Rovelli, {\it Barrett-Crane model from a Boulatov-Ooguri field theory over a homogeneous space},
\newblock Nucl. Phys. \textbf{B574} (2000): 785-806
  [\href{http://arxiv.org/abs/hep-th/9907154}{arXiv: hep-th/9907154}].

\bibitem{Carlip:2014bfa}
S.~Carlip, {\it A Note on Black Hole Entropy in Loop Quantum Gravity},
\newblock Class. Quant. Grav. \textbf{32} (2015): 155009
  [\href{http://arxiv.org/abs/1410.5763}{arXiv: 1410.5763}].

\bibitem{Liu:2017bfk}
Hongguang Liu and Karim Noui, {\it Gravity as an SU(1,1) gauge theory in four dimensions},
\newblock ArXiv e-prints (2017):
  [\href{http://arxiv.org/abs/1702.06793}{arXiv: 1702.06793}].

\bibitem{Carlip:1994gy}
Steven Carlip, {\it The Statistical Mechanics of the (2+1)-Dimensional Black Hole},
\newblock Phys. Rev. \textbf{D51} (1995): 632-637
  [\href{http://arxiv.org/abs/gr-qc/9409052}{arXiv: gr-qc/9409052}].

\bibitem{Barenz:2016nzn}
Manuel B\"{a}renz and John Barrett, {\it Dichromatic state sum models for four-manifolds from pivotal functors},
\newblock ArXiv e-prints (2016):
  [\href{http://arxiv.org/abs/1601.03580}{arXiv: 1601.03580}].

\end{thebibliography}
\end{document}